\documentclass[english,twocolumn,aps]{revtex4}
\usepackage[T1]{fontenc}
\usepackage[latin9]{inputenc}
\usepackage{amsmath}
\usepackage{graphicx}
\usepackage{amssymb}

\usepackage{babel}

\begin{document}


\title{Exploiting Kerr Cross Non-linearity in Circuit Quantum Electrodynamics
for Non-demolition Measurements}

\author{Shwetank Kumar and David P. DiVincenzo}


\affiliation{IBM T. J. Watson Research Center, Yorktown Heights, NY 10598 USA}
\begin{abstract}
We propose a scheme for dispersive readout of stored energy in one
mode of a nonlinear superconducting microwave ring resonator by detection
of the frequency shift of a second mode coupled to the first via a
Kerr nonlinearity. Symmetry is used to enhance the device responsivity
while minimizing self nonlinearity of each mode. Assessment of the
signal to noise ratio indicates that the scheme will function at the
single photon level, allowing quantum non-demolition measurement of
the photon number state of one mode. Experimental data from a simplified
version of the device demonstrating the principle of operation are
presented.

PACS: 84.40.Dc, 85.25.-j, 85.25.Am
\end{abstract}
\maketitle
High quality factor superconducting microwave resonators are versatile
devices that promise to form an essential part of a scalable quantum
computing architecture. They have been used to readout superconducting
qubits (flux \cite{Lee2007}, charge \cite{Wallraff2005}, transmon
\cite{Koch2007}), and to provide a long-lived memory for qubit states
via {}``parking'' \cite{Koch2006}. Integrated with SQUIDs they
have been used to implement high-sensitivity parametric amplifiers
\cite{Abdo2009,Beltran2008}, create tunable resonators for the
selective coupling of multiple qubits \cite{Laloy2008,Sandberg2008,Wallquist2006},
and study dephasing due to Kerr coupling between resonator
modes \cite{Suchoi2009}. In two-qubit systems, resonators function
as a cavity bus to couple far separated qubits \cite{Majer2007}.
Coupled to polar molecules \cite{Andre2006} and nanomechanical resonators
\cite{Teufel2008} they have been proposed for sideband cooling experiments and
for the construction of exotic superpositions of states. Theoretical
proposals have been advanced for construction of non-classical states
by using the Kerr coupling between two nanomechanical resonators \cite{Seminao2009}
and for dispersive readout of number states of these resonators using
the anharmonic coupling between two beam-bending modes \cite{Santamore2004}.
Since superconducting resonators may also be used as quantum memory
elements \cite{Pritchett2005}, and as subcomponents of a qubit \cite{Koch2007}
or a quantum repeater \cite{DiVincenzo2008}, the ability to directly
measure their quantum state is invaluable.

In this paper we present a superconducting circuit that realizes a
non-demolition measurement of the energy stored in a resonator
mode. Realistic simulations and calculations based on measured noise
data \cite{Gao2008, ringmod} indicate that single-photon sensitivity will
be achievable. Our proposal uses Kerr coupling between two modes of the resonator to measure the shift in the frequency of one mode as the other mode is populated. The non-demolition condition \cite{Thorne1978}, that the state projected into by the measurement be an eigenstate of the system, is enforced in a precise way by parity symmetry in our device, in which an odd-parity state is measured by an even-parity mode. Impressive progress on non-demolition measurements in superconducting devices, with single-quantum sensitivity, have been reported recently in various qubit-resonator systems, with either the qubit \cite{Lupascu2007} or the resonator photons \cite{Schuster2007,Wang2008,Guerlin2007} as the measured quanta.  Our proposal is distinguished from this previous work in two ways: First, the non-demolition condition, which is generally violated by small off-resonant terms (as in \cite{Schuster2007}), is enforced {\em exactly} by symmetry in our circuit.  Second, there is no qubit involved; both our measured and measuring system are weakly anharmonic resonator modes.  A measurement scheme exploiting the Kerr mechanism between two such modes has also been invoked in \cite{Helmer2009}, but only as part of a scheme for measuring {\em itinerant} photons. The parameters necessary for achieving single-photon sensitivity in our circuit are readily accessible using current superconducting technology, unlike the situation in traditional quantum optics \cite{Xiao2008}.

\begin{figure}
\noindent \begin{centering}
\includegraphics[scale=0.4]{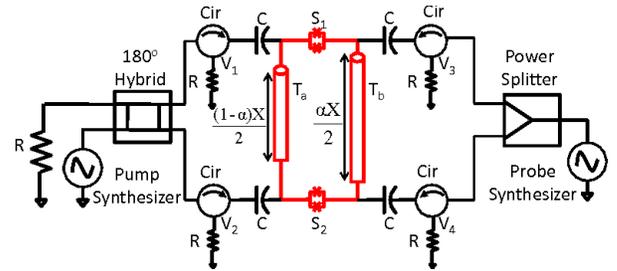}
\par\end{centering}

\caption{Proposed circuit for implementation of 
nondemolition measurement using the Kerr effect. 
The superconducting ring resonator (red) is made from two transmission lines $T_{a,b}$ of unequal lengths (as indicated) joined at the ends by nonlinear elements realized by nominally identical DC SQUIDs $S_{1,2}$. The 180-degree hybrid causes the pump to couple only to ring modes with an odd voltage profile with respect to the horizontal midline of the device; the power splitter causes the probe to couple only to even modes.}

\centering{}\label{fig:Circuit}
\end{figure}

Figure \ref{fig:Circuit} shows a schematic for the proposed
circuit.
The ring resonator circumference is $X=16.00$ mm. The DC SQUID two-terminal
response 
is that of a nonlinear inductance, described
in leading order by 
\begin{equation}
L_{sq}=L_0+L_1I^{2}.\label{nlind}
\end{equation}
We use 
$L_0=0.19$
nH and 
$L_1=32$ H/$\text{A}^{2}$
\cite{Sandberg2008}, corresponding
to an unbiased DC SQUID with critical current $I_{c1}=0.85$
$\mu$A per junction; inductance parameters can be tuned with flux
biasing.  For the transmission line segments T$_{a,b}$, we assume an inductance
and capacitance per unit length of ${\cal L}=0.45\text{ pH}/\mu\text{m}$,
${\cal C}=0.16\text{ \text{fF}}/\mu\text{m}$,
appropriate for coplanar waveguide (CPW) geometry with 3 $\mu$m
center strip and 2 $\mu$m slots. The loss tangent and the effective
dielectric constant are $\tan\delta=10^{-5}$ and $\epsilon_{eff}=6.45$,
typical for a high purity silicon substrate \cite{Connell2008}. The ring resonator is excited using two frequency synthesizers (pump and probe) through four capacitors with $C=10$ fF.

Eq.~(\ref{nlind}) is modified if we take account of the effect of another flux, that threading the large ring resonator loop, $\Phi_{ring}$.  This effect is very small because the self-inductance of the full ring $L_t$ is quite large. $\Phi_{ring}$ slightly changes the DC operating point for the superconducting phase around the ring, thereby changing the phase values around which the AC dynamics occurs.  The full expression for the effective inductance including the effect of $\Phi_{ring}$ is:
\begin{eqnarray}
L&=&L_0\left[1+\left({L_0\over L_t}\right)^2 z^2\right]+{2\pi\over\Phi_0}{\sqrt{2}L_0\over L_t} zL_0^2I+\nonumber\\&&L_1\left[1+9\left({L_0\over L_t}\right)^2z^2\right]I^2.
\end{eqnarray}
Here $z=2\pi\Phi_{ring}/\Phi_0$. We estimate $L_t=16{\rm mm}\times0.45$pH/$\mu$m=7.2nH. For this value of $L_t$ the modification of the leading term, and of the $I^2$ term, is negligible.  A new term proportional to $I$ is present for $z\neq 0$, but its contribution will vanish in the rotating wave approximation used below.  

The fundamental even and odd resonant modes, degenerate
in the ideal ring, are split in frequency by coupling to the SQUID
inductances. In the linear regime the even ($e$) and odd ($o$) mode
frequencies are
\begin{equation}
f_{e,o}=\bar{f}-f_{J}(1\pm\cos(\pi\alpha))
\label{eq:resfreq}
\end{equation}

where $\bar{f}=\frac{1}{X\sqrt{{\cal LC}}}$ and $f_{J}=\bar{f}^2L_0/Z_0$ with $Z_{0}=\sqrt{{\cal L}/{\cal C}}$.
It will be experimentally convenient to explore cross-Kerr effects in two modes that
are close in frequency; our odd and even modes can be split by any
desired amount (from 0 to hundreds of MHz) by choosing $\alpha$,
determined by the relative lengths of $T_{a}$ and $T_{b}$ as shown in Fig. 1.
We write a classical Hamiltonian for these two modes. The harmonic part is exactly as in \cite{Wallquist2006}, while the additional anharmonic parts of the energy go like ${1\over 4} L_1(I_e\pm I_o)^4$ ($\pm$ for the SQUIDs $S_1$ and $S_2$ respectively), where $I_{e/o}$ are the mode currents at the location of the SQUIDs.

In terms of the standard conjugate variables $n$ and $\phi$ \cite{Wallquist2006} we obtain a Hamiltonian describing two harmonic modes, with additional purely quartic nonlinear terms:
\begin{widetext}
\begin{eqnarray}
H & = & E_{Ce}n_{e}^{2}+E_{Le}\phi_{e}^{2}+E_{Co}n_{o}^{2}+E_{Lo}\phi_{o}^{2}+\frac{16\pi^{2}L_1}{\Phi_{0}^{4}}\left[E_{Le}^{4}\phi_{e}^{4}\cos^{4}\beta+E_{Lo}^{4}\phi_{o}^{4}\sin^{4}\beta+6E_{Le}^{2}E_{Lo}^{2}\phi_{e}^{2}\phi_{o}^{2}\sin^{2}\beta\cos^{2}\beta\right]
\label{eq:Enl}
\end{eqnarray}
where $\beta=\pi\alpha/2$, and the parameters $E_{(C,L)(o,e)}$,
the electric and magnetic energies in the modes, are as in \cite{Wallquist2006}.
Our symmetrical geometry causes many possible additional nonlinear
terms, e.g., ones proportional to $\phi_{e}\phi_{o}^{3}$, to be absent
\cite{ringmod}).  Eq.
(\ref{eq:Enl}) can be quantized in the standard way \cite{Wallquist2006};
in the rotating wave approximation (RWA) we obtain
\begin{eqnarray}
H^{RW} & = & \hbar\omega_{e}a_{e}^{\dagger}a_{e}+\hbar\omega_{o}a_{o}^{\dagger}a_{o}+\frac{4\pi^{2}L_1}{\Phi_{0}^{4}}\text{ }\left\{ 12\sqrt{E_{Le}^{3}E_{Lo}^{3}E_{Ce}E_{Co}}\left[2a_{e}^{\dagger}a_{e}a_{o}^{\dagger}a_{o}-a_{o}^{\dagger}a_{o}-a_{e}^{\dagger}a_{e}\right]\sin^{2}\beta\cos^{2}\beta+\right.\nonumber \\
 &  & E_{Le}^{3}E_{Ce}\left[(a_{e}^{\dagger}a_{e})^{2}-6a_{e}^{\dagger}a_{e}\right]\cos^{4}\beta+E_{Lo}^{3}E_{Co}\left[(a_{o}^{\dagger}a_{o})^{2}-6a_{o}^{\dagger}a_{o}\right]\sin^{4}\beta\biggr\}.
\label{kerr}
\end{eqnarray}
\end{widetext}
Here $\hbar\omega_{i}=2\sqrt{E_{Li}E_{Ci}}$. Note the cross-Kerr
type term proportional to $a_{e}^{\dagger}a_{e}a_{o}^{\dagger}a_{o}$.
Terms in the rotating frame that are proportional to
$a_{e}a_{e}a_{o}^{\dagger}a_{o}^{\dagger}e^{i(2\omega_{e}-2\omega_{o})t}$
(in the interaction picture) are dropped from the Hamiltonian. This
term and other similar ones are not so rapidly time varying if $f_{e}\approx f_{o}$.
For our circuit $f_{e}-f_{o}\le50$ MHz,
which means that this additional term could definitely not be
ignored in nsec-scale pulsed experiments \cite{Sandberg2008}. However,
here we only consider the steady state circuit response.
Further, we can, without 
changing any other aspect of
the model, 
raise the even-odd splitting with simple
stub-tuning techniques \cite{Chirolli2010}. Thus, these terms need not be a
concern in
the discussion of the quantum aspects of our proposal.

\begin{figure}
\noindent \begin{centering}
\includegraphics[scale=0.16]{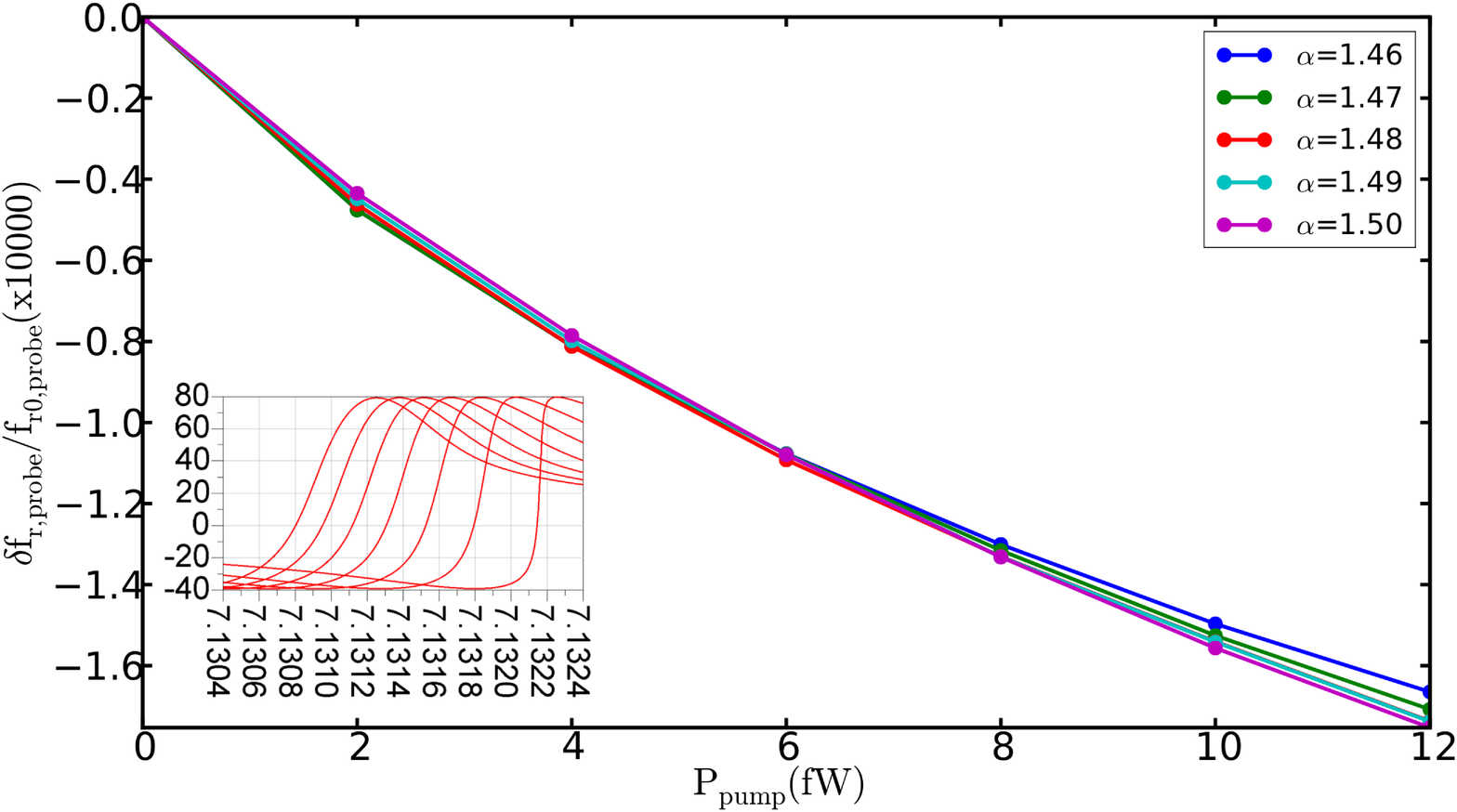}
\par\end{centering}

\caption{Simulated device response near $\alpha=3/2$, where $\alpha$ fixes the relative lengths of $T_{a}$ and $T_{b}$ (see Fig. 1). Inset:
phase response of $M_{1}$ as pump power $P_{pump}$ is swept from
0 to 12 fW in steps of 2 fW for $\alpha=3/2$.
$f_{r0,probe}$ = 7.1322 GHz is the probe resonance
frequency in the absence of any pump excitation, and $\delta f_{r,probe}$ is the shift of the probe resonant response under probe excitation.
}

\centering{}\label{fig:Response}
\end{figure}

The circuit has been simulated using the Agilent Advanced Design System(ADS) \cite{ADS} for values of $\alpha$ near 3/2, where
the pump and probe nonlinearities are comparable (Eq. (\ref{eq:Enl})).
The resonance frequencies of the circuit
are determined using linear analysis in which the circuit is excited
by sweeping the frequency of the pump and probe synthesizers and measuring
the reflected response at each of the four ports. The reflected wave amplitudes are measured as voltage drops $V_{1-4}$
across the output resistors $R=50\text{ }\Omega$ using the circulators to distinguish between the incident and reflected waves.  The even and odd
response is extracted as $M_{1,2}=V_{1}\pm V_{2}$ ($M_{3,4}=V_{3}\pm V_{4}$ could also be used).
These low-power resonance frequencies match very
well with Eq. (\ref{eq:resfreq}).

The finite-power circuit response is simulated using the Harmonic
Balance (HB) tool in ADS \cite{ADS}.  This classical tool is suitable for extracting parameters for our simple quantum description, and for confirming that all significant aspects of the system's response is captured by our Hamiltonian. In these simulations, the pump synthesizer is
set to the pump resonance frequency and the probe synthesizer frequency
is swept to extract the probe resonance feature for different pump
powers. The results of the HB simulation, for circuits
with different values of $\alpha$ near 3/2, are shown in Fig.
\ref{fig:Response}. 

From this
data we extract the fractional shift of the probe frequency
as a function of pump power.
A cross Kerr nonlinearity is clearly seen, as expected from Eqs.
(\ref{eq:Enl},\ref{kerr}); at small pump powers, the frequency of the resonant
response seen by the probe shifts linearly with pump power.  The probe power is
set to $P_{probe}=5.3$ fW so that we obtain a signal to noise ratio sufficient
to detect single photons.  At this probe power level the probe resonant response
shows significant asymmetry due to the self-nonlinearity of the mode (Eq.
(\ref{kerr})), but it is still below the power where bifurcation takes place. 
At the highest pump powers it is evident that the circuit exhibits higher-order
nonlinearities not described by Eq. (\ref{kerr}).  But the model is very good at
the very low pump powers that are relevant for describing the behavior of the
measurement when only a few photons are in the pump mode.  

We now calculate the expected signal to noise ratio, showing that single-photon detection is indeed possible.
The pump power at resonance required to maintain
one-photon circulating in the resonator in this mode
is $P_{photon}=2\pi hf_{r,pump}^{2}/Q_{r,pump}$, where $f_{r,pump}$, $Q_{r,pump}$ are
its resonance frequency and quality factor, respectively. The quality factor of the pump mode of our circuit is coupling limited to $\sim$3,000 and it has a resonance frequency of 7.12 GHz.
Then $P_{photon}\approx 0.07$ fW.  By performing HB simulations of the phase response at $P_{pump}=0$ and 0.07 fW, we read off that the expected single-photon phase
signal is $\theta_{sig}=19^{\circ}$.

This signal level is to be compared with the rms deviations of the detected
phase due to several sources of noise.  Noise due to amplification is
fundamentally no less than one-half a photon of power per signal mode, but in
practice is determined by an amplifier temperature $T_N$.  The rms magnitude of
the variations in the detected phase due to this noise is \cite{ringmod}
\begin{equation}
\sigma_\theta^{amp.}={4k_BT_n\Delta\nu\over P_{probe}}.
\end{equation}
For state-of-the-art HEMT amplifiers, $T_N\approx 4$K \cite{ringmod}.  The
optimal bandwidth is set by the probe linewidth, and is roughly estimated as
$\Delta\nu\approx f_{r,probe}/Q_{r,probe}$. Here $Q_{r,probe}$ is the probe
quality factor, which is also coupling limited to 3000, and $f_{r,probe}\approx
7.1$GHz; thus, $\Delta\nu\approx 1$MHz.  This gives
\begin{eqnarray}
\sigma_\theta^{amp.}&\approx& 2.7^{\circ}\\
{\mbox{SNR}}=\theta_{sig}/\sigma_\theta&\approx& 7.1\,\,\,\mbox{(single photon)}.
\end{eqnarray}

We need to confirm that amplifier noise is indeed the determining limitation on the signal to noise ratio.  Since phase noise due to low-frequency fluctuations in the resonant frequency is a dominant source of noise in experiments performed at higher power levels, we examine this mechanism in detail. To estimate the effects of this noise, we extrapolate results from \cite{Gao2007}.
The fractional frequency noise of a CPW resonator similar to the one we have assumed is measured
there to be $S_{\delta f}(\nu_{0},\text{ }P_{0})/f_{r}^{2}=10^{-18}$/Hz
at bandwidth $\nu_{0}=1$ kHz. Here $P_{0}=-60\mbox{ dBm}=1\mbox{ nW}$
corresponds to what we call probe power. Frequency noise power can
be converted to phase noise using $S_{\theta}(\nu_{0},\text{ }P_{0})=4Q_{r,probe}^{2}S_{\delta f}(\nu_{0})/f_{r}^{2}=3.6\times10^{-11}\text{ rad}^{2}/\text{Hz}$.  
The phase noise in the bandwidth $\Delta\nu$ of our system can be estimated as
$S_{\theta}(\Delta\nu,\text{
}P_{0})=\left(\Delta\nu/\nu_0\right)^{-1/2}S_{\theta}(\nu_{0},\text{ }P_{0}) =
1.1\times10^{-12}\text{ }\text{ rad}^{2}/\text{Hz}$. The resonator phase noise
is drive power dependent and needs to be scaled for the probe power used in our
simulation.
For a coupling-limited resonator, the internal power is
given by $P=P_{probe}Q_{r,probe}$; then the phase noise power for our parameters
can be estimated to be $S_{\theta}(\nu,\text{ }P)=\left(P/P_0\right)^{-1/2}S_{\theta}(\nu,\text{ }P_{0})=8.3\times 10^{-12}$
$\text{rad}^{2}/\text{Hz}$. Finally, the phase measurement error is:
\begin{equation}
\sigma_{\theta}^{ph.}=\sqrt{S_{\theta}(\nu,\text{ }P)/\tau}=0.18^{\circ}
\end{equation}
Thus,
frequency fluctuations make an insignificant contribution to the SNR.  We are
unaware of any other mechanisms that are likely to contribute at this level. 

\begin{figure}[b]
\noindent \begin{centering}
\includegraphics[scale=0.2]{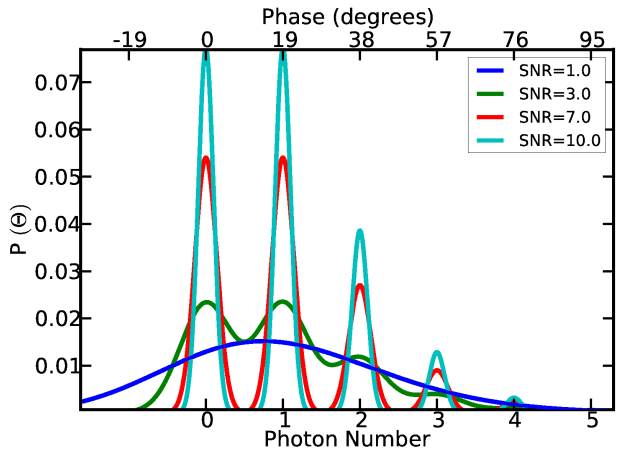}
\par\end{centering}

\caption{Predicted histogram of phase measurement outcomes $\theta$ (measured
relative to the zero pump power phase) for different measurement signal
to noise ratios (SNR) for the device in Fig. 1. Measurement
is simulated for a coherent state with average photon number $n_{av}=1$.
The quantum nature of the radiation field is not evident for 
SNR=1, but for the predicted values of SNR=7.1 clearly resolved
photon-number peaks will be seen.
}

\centering{}\label{fig:Histogram}
\end{figure}

Figure \ref{fig:Histogram} shows that our value of SNR=7.1
will be sufficient to discern the quantum nature of the resonator
state. We plot the probability density function for repeated measurements
of the Kerr-induced phase shift $\theta$, in the probe microwave signal. This is given by a sum over the possible number-state outcomes, convolved with a function describing the additive noise in the measurement: $P(\theta)=\sum_{n=0}^{\infty}P(\theta|n)P(n)$, where $P(n)$ is the probability of projecting the coherent state into the $n^{th}$ number state given by the Poisson distribution ($P(n)=\frac{1}{en!}$ for $n_{av}=1$), and $P(\theta|n)$ is the probability of measuring the $\theta$ given that the mode has been projected into photon number $n$. We assume Gaussian additive noise $P(\theta|n)=\frac{1}{\sigma_{\theta}\sqrt{2\pi}}\exp\left(-\frac{(\theta-n\theta_{sig})^{2}}{2\sigma_{\theta}^{2}}\right)$
where $\sigma_{\theta}$ and $\theta_{sig}$ are estimated above.
The $x$-axis of Fig. 3 is marked with
the phase shift $n\theta_{sig}$ corresponding to the different photon numbers $n$
in a noiseless measurement. We see that even for SNR$\sim 3$,
much less than the estimated value, our circuit should readily
detect quantization of the coherent state. Various nonidealities, such as non-identical SQUID critical currents ($I_c$), introduce extra nonlinear terms which reduce our response. However,
our simulation shows that expected $\pm 5$\% variations of $I_c$ reduce the response only slightly.  In fact, there is considerable scope for improving the SNR by simultaneously measuring amplitude and phase response \cite{Gao2007} and by carefully designing the resonator geometry to reduce phase noise \cite{Gao2008}.
%
%

\begin{figure}[t]
\noindent \begin{centering}
\includegraphics[scale=0.34]{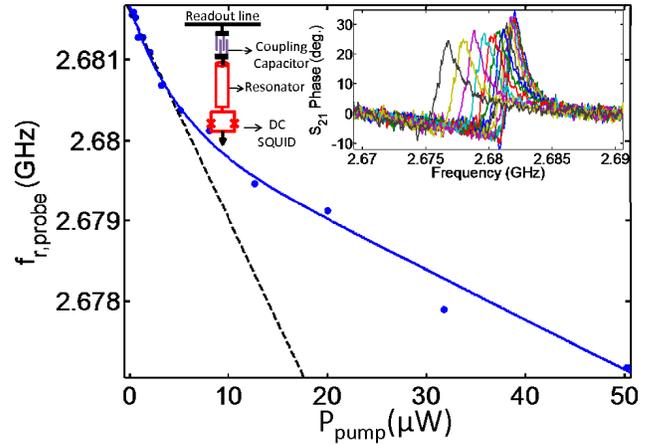}
\par\end{centering}

\caption{Experimentally measured cross-Kerr response of simple SQUID-tuned
resonator. The circuit schematic is much simpler than the optimized
one above (Fig. 1). A CPW resonator with 3 $\mu$m center strip and 2 $\mu$m slots was used. The DC SQUID loop had an area of 5$\times$5 $\mu$m$^2$ with each junction having $I_c=1.2$ $\mu$A. The SQUID was not flux biased for data in this figure.
Readout line was used for pump
excitation as well as measuring the forward scattering parameter $S_{21}$
for the probe signal. Inset shows resonant phase response of the probe
for different values of the pump power. Probe resonance frequencies
are plotted in the main plot; the blue line is guide to the eye. The
initial linear behavior of this curve at low pump power (slope indicated
by black dashed line) is a manifestation of the cross-Kerr effect.}
\centering{}\label{fig:ExptResponse}
\end{figure}

We have performed an experiment which shows that the basic effect predicted here is readily seen,
even in a device that is not carefully designed to optimize the cross-Kerr
effect. The circuit is a
linear niobium transmission line resonator
with a quarter-wave geometry, capacitively coupled to a readout line
at the top (Fig. \ref{fig:ExptResponse}) and shorted to ground through a nonlinear inductor (an
Al/AlO$_{x}$/Al DC SQUID) at the bottom end. This device was measured in a He-3 cryostat at 360 mK.

It has fundamental (probe) and third harmonic (pump) resonances at 2.68 GHz and 8.05 GHz, respectively.
The probe mode has a coupling-limited quality factor $Q_{r}\simeq2000$.
The upper right inset of Fig. \ref{fig:ExptResponse} shows the phase response of the device
for different pump drives. From these we obtain Kerr shift
of the probe as a function of the pump mode power. The resonance frequency is evaluated as the frequency at which the derivative of phase response with respect to the readout frequency is maximum. The probe power is held constant at -115 dBm.
A detectable Kerr shift 
is observable for pump power as low as -70 dBm. A decrease in resonator quality factor similar to \cite{Suchoi2009} is observed as well. While our choice of materials and operating temperatures allows us to measure these nonlinearity-induced effects at relatively low pump powers, accessing the quantum regime requires carefully optimised design proposed in Fig. \ref{fig:Circuit}.


To summarize, we have presented and analyzed a circuit for dispersive readout
of energy stored in one mode of a superconducting microwave resonator
by measuring the frequency shift of another mode coupled
to the first through a cross-Kerr nonlinearity.
The proposed device is sensitive
enough to operate at the single photon energy level, and it can
be used to perform a quantum non-demolition measurement of the photon
number state of the resonator. The basic effect is readily seen in
a simple experiment using an unoptimized version of the device. Many
other uses can be envisioned for this device; for example, modulation
of the SQUID flux bias at $f_{e}-f_{o}$ will implement a beam splitter
between the two modes~\cite{Chirolli2010}.

We thank Dr. John Kirtley and the IBM experimental quantum computing team
for useful discussions. We thank the IBM Research Communication
Technologies Department for use of Agilent Design Systems and are grateful
to M.B. Rothwell and G.A. Keefe for the devices. DD is grateful to the Institute
for Theoretical Physics at the University of Amsterdam for its hospitality
during the writing of this paper, and for support from an ESF Exchange
Grant under the INSTANS program.


\begin{thebibliography}{29}
\bibitem{Lee2007}J.C. Lee \emph{et al.}, Phys. Rev. B 75, 144505 (2007); A.A. Abdumalikov \emph{et al.}, Phys. Rev. B 78, 180502(R) (2008)

\bibitem{Wallraff2005}A. Wallraff \emph{et al.}, Phys. Rev. Lett. 95, 060501 (2005)

\bibitem{Koch2007}J. Koch \emph{et al.}, Phys. Rev. A 76, 042319 (2007)

\bibitem{Koch2006}R.H. Koch \emph{et al.}, Phys. Rev. Lett. 96, 127001 (2006)

\bibitem{Abdo2009}B. Abdo \emph{et al.}, Euro. Phys. Lett. 85, 68001 (2009).

\bibitem{Beltran2008}M.A. Castellanos-Beltran \emph{et al.}, Nat. Phys.,
4, 929 (2008).

\bibitem{Laloy2008}A. Palacios-Laloy \emph{et al.}, J. of Low
Temp. Phys., Vol. 151, 3, pp 1034-1042 (2008)

\bibitem{Wallquist2006}M. Wallquist, V.S.Shumeiko, G. Wendin, Phys. Rev. B, 74, 224506 (2006)

\bibitem{Sandberg2008}M. Sandberg \emph{et al.}, Appl. Phys. Lett. 92, 203501 (2008)

\bibitem{Suchoi2009}O. Suchoi \emph{et al.}, arXiv:0901:3110

\bibitem{Majer2007}J. Majer \emph{et al.}, Nature, 449, 443 (2007)

\bibitem{Andre2006}A. Andr$\acute{\text{e}}$ \emph{et al.}, Nature Physics, 2,
636 (2006)

\bibitem{Teufel2008}J.D. Teufel \emph{et al.}, Phys. Rev. Lett., 101, 197203
(2008)

\bibitem{Seminao2009}F.L. Semi$\bar{\text{a}}$o , K. Furuya, G.J. Milburn,
arXiv:0808.0743

\bibitem{Santamore2004}D. H. Santamore, A.C. Doherty, and M.C. Cross, Phys. Rev. B 70, 144301 (2004)

\bibitem{Pritchett2005}E.J. Pritchett and M.R. Geller, Phys. Rev. A 72, 010301R
(2005)

\bibitem{DiVincenzo2008}D.P. DiVincenzo, P.C.D. Hobbs, S. Kumar, Internal Communication
(2008)

\bibitem{ringmod}N. Bergeal \emph{et al.}, arXiv preprints arXiv:0805.3452,
arXiv:0912.3407.

\bibitem{Gao2008}J. Gao \emph{et al.}, Appl. Phys. Lett.
92, 212504 (2008)

\bibitem{Thorne1978}K. S. Thorne \emph{et al.}, Phys. Rev. Lett., 40, 667-671
(1978).

\bibitem{Lupascu2007} A. Lupascu \emph{et al.}, Nature Physics 3, 119 (2007)

\bibitem{Schuster2007}D. I. Schuster,\emph{et al.}, Nature, 445, 515 (2007)

\bibitem{Wang2008}H. Wang \emph{et al.}, Phys. Rev. Lett. 101, 240401 (2008)

\bibitem{Guerlin2007} C. Guerlin \emph{et al.}, Nature 448, 889-893 (2007)

\bibitem{Helmer2009} F. Helmer \emph{et al.}, Phys. Rev. A 79 (5), 052115 (2009)

\bibitem{Xiao2008} Y. Xiao \emph{et al.}, Optics Express 21462, Vol. 16, No. 26
(2008)

\bibitem{Connell2008} A. D. O'Connell \emph{et al.}, Appl. Phys. Lett. 92,
112903 (2008)

\bibitem{Chirolli2010} L. Chirolli \emph{et al.}, arXiv
preprint arXiv:1002.1394.

\bibitem{ADS}Agilent Advanced Design System URL:
http://eesof.tm.agilent.com/products/adsmain.html

\bibitem{Gao2007}J. Gao \emph{et al.}, Appl. Phys. Lett. 90, 102507
(2007)



\end{thebibliography}
\end{document}